%% file: jstat2_rev.tex
\documentclass[12pt]{iopart}

\usepackage{iopams}  
\usepackage[dvips,dvipdfm]{graphicx}

\input{commands}

\begin{document}

\title[Analytic properties of random energy models II]
{On analyticity with respect to the replica number in 
random energy models II: zeros on the complex plane}
\author{Kenzo Ogure$^1$ and Yoshiyuki Kabashima$^2$}

\address{$^1$
Department of Nuclear Engineering, 
Faculty of Engineering, Kyoto University, 
Kyoto 606-8501, Japan \\
$^2$Department of Computational Intelligence and Systems Science, 
Tokyo Institute of Technology, Yokohama 226-8502, Japan\\}
\ead{ogure@nucleng.kyoto-u.ac.jp$^1$, kaba@dis.titech.ac.jp$^2$}

\begin{abstract}
We characterize the breaking of analyticity with respect to 
the replica number which occurs in random energy models via the 
complex zeros of the moment of the partition function. 
We perturbatively evaluate the zeros in the vicinity of 
the transition point based on an exact expression of
the moment of the partition function utilizing 
the steepest descent method, and examine an asymptotic form of the 
locus of the zeros as the system size tends to infinity. 
The incident angle of this locus indicates that analyticity 
breaking is analogous to a phase transition of the second order. 
We also evaluate the number of zeros utilizing the argument principle
of complex analysis. 
The actual number of zeros calculated numerically
for systems of finite size agrees fairly well  
with the analytical results. 
\end{abstract}

\maketitle

\section{Introduction}
The replica method (RM) is a powerful tool in statistical mechanics
for the analysis of disordered systems 
\cite{Dotzenko2001} . 
In general, the objective of RM is to 
evaluate the generating function 
\begin{eqnarray}
\phi_N(n)=\frac{1}{N} \log \left \langle Z^n \right \rangle, 
\label{cumulant}
\end{eqnarray}
for real {\em replica numbers} $n \in \mR$ (or the complex field $\mC$). 
Here, $Z$ is the partition function, 
$N$ represents the system size and 
$\left \langle \cdots \right \rangle$ denotes
the configurational average with respect 
to the external randomness which governs the objective system. 
Direct evaluation of eq. (\ref{cumulant})
is generally difficult. However, eq. (\ref{cumulant})
can often be evaluated for natural numbers $n=1,2,\ldots \in \mN$ 
in the thermodynamic limit as $N \to \infty$. 
Therefore, in RM, one usually evaluates 
\begin{eqnarray}
\phi(n)=\lim_{N \to \infty} \phi_N(n) 
\label{phin}
\end{eqnarray}
for $n \in \mN$, in order to first obtain an analytic expression 
for $\phi(n)$ and then analytically continues
this expression to $n \in \mR$ (or $\mC$). 

There are two known possible problems with  this procedure. 
The first problem is multiple possible alternatives for 
analytic continuation \cite{vanHemmenPalmer1979}. 
Even if all values of $\left \langle Z^n \right \rangle$ are
provided for $n \in \mN$, analytic continuation of 
$\phi_N(n)$ from $n \in \mN$ to $n \in \mR$ (or $\mC$) is not
uniquely determined. 
van Hemmen and Palmer conjectured that 
this may be the origin of the failure of the replica symmetric
(RS) solution in the low temperature phase of the 
Sherrington-Kirkpatrick (SK) model \cite{SK}. 
The other issue is the possible breakdown of analyticity of $\phi(n)$. 
Although analyticity of $\left \langle Z^n \right \rangle$ 
with respect to $n$ is 
generally guaranteed as long as $N$ is finite, 
$\phi(n)=\lim_{N \to \infty}\phi_N(n)$ may fail to be analytic. 
This implies that if such a breaking of analyticity occurs at $n=n_c < 1$,
then continuing the expression analytically from $n \in \mN$ to $n \in \mR$ 
will lead to an incorrect solution for $n$ in the range of $n < n_c$. 

In \cite{KO1,KO2}, the authors developed an exact expression of the 
moment $\left<Z^n\right>$ for discrete versions of random energy 
models (DREMs) \cite{Mou1,Mou2}. The expression is valid for 
$n \in \mC$ and is useful for handling systems of finite size. 
Utilizing this expression, it can be shown that analyticity breaking 
of $\phi(n)$ actually occurs 
at a certain critical replica number $n=n_c < 1$ in the low 
temperature phase of DREMs. 
The uniqueness of the analytic continuation from $n \in \mN$ 
to $n \in \mC$ is guaranteed for DREMs. 
This means that the analyticity breaking 
of $\phi(n)$ with respect to $n$ is the origin of the 
one step replica symmetry breaking (1RSB) which 
is observed for DREMs, and thus we are motivated to further 
explore its mathematical structure. 

This paper is written with this motivation in mind. 
Regarding $\log Z$ as the energy of the external 
randomness, the moment $\left \langle Z^n \right \rangle$ 
and generating function $\phi(n)$ are
formally analogous to the partition function and free energy 
of non-random systems. This analogy leads us to 
characterize analyticity breaking in terms of the distribution 
of zeros of $\left \langle Z^n \right \rangle$ on the complex $n$ plane, 
following the argument by Lee and Yang \cite{Yan1,Yan2,Gro1,Gro2}. 
Let us suppose that a partition function of a finite size system 
is expressed as a function of a certain control parameter $\zeta$. 
As long as $\zeta$ is real, the partition function 
does not vanish. However, there can exist zero points, 
which are generally termed the Lee-Yang zeros, throughout the complex $\zeta$ plane. 
In general, the Lee-Yang zeros are apart from the real axis as long 
as the system size $N$ is finite. However, when a certain phase transition 
occurs at a critical parameter value $\zeta_c \in \mR$, the 
distribution of the zeros becomes dense and approaches 
$\zeta_c$ as $N \to \infty$.  The way in which this occurs 
characterizes the type of the phase transition. 
The main purpose of this paper is to apply such a description 
to the analyticity breaking of $\phi(n)$ by evaluating the zeros 
of $\left \langle Z^n \right \rangle$
on the complex plane of the replica number $n$.

This paper is organized as follows. In section \ref{lyzero}, 
we briefly discuss the Lee-Yang zeros, 
utilizing a simple model.  
In section \ref{sec:replica}, which is the main part of 
this paper, we apply the Lee-Yang approach to examine 
the analyticity breaking of $\left \langle Z^n \right \rangle $
of DREM. The exact expression for 
$\left \langle Z^n \right \rangle$ developed in 
\cite{KO1,KO2} is utilized to 
perturbatively evaluate the zeros in the vicinity of 
the transition point employing the steepest descent method. 
This shows that the the distance of the closest zero 
to the real axis decays as $O(N^{-1/2})$
when the system size $N$ is large. 
In the thermodynamic limit as $N \to \infty$, 
the incident angle of the locus to the real axis 
converges to $\pi/4$, indicating that analyticity 
breaking is analogous to a phase transition of the second order. 
However, it is also shown that the angle between the real axis 
and the line connecting the transition point with the $k$th 
zero from the real axis has a finite positive correction 
from $\pi/4$ independently of $N$ as long as $k(=1,2,\ldots)$ is finite. 
The complex zeros are also numerically evaluated for several 
system sizes based on the expression, which 
is consistent with 
the asymptotic form to a reasonable precision. 
The final section is devoted to a summary.

\section{Complex zeros and analyticity}\label{lyzero}
\subsection{Brief review of Lee-Yang zeros}\label{ordinary}
Partition functions of discrete systems of finite size are, 
in general, analytic with respect to their parameters because 
they are a summation of exponents of the Hamiltonian over all possible (but still 
a finite number of) states.  
In addition, they do not have zeros on the real axis for the same reason.  

A theorem by Weierstrass \cite{Alf},
\begin{quote}
``{\it There exists an entire function with arbitrarily prescribed 
zeros $a_n$ provided that, in the case of infinitely many zeros, 
$a_n\rightarrow\infty$.  Every entire function with these 
and no other zeros can be written in the form
  \begin{eqnarray}
    \label{eq:wei}
    f(z)=z^me^{g(z)}\prod_{n=1}^{\infty}(1-\frac{z}{a_n})
    e^{\frac{z}{a_n}+\frac{1}{2}(\frac{z}{a_n})^2+\cdots+\frac{1}{m_n}(\frac{z}{a_n})^{m_n}}
  \end{eqnarray}
where the product is taken over all $a_n\neq 0$, the $m_n$ are certain integers, and $g(z)$ is an entire function.}",
\end{quote}
indicates that a partition function $Z(\zeta)$ of 
a finite system can generally be expanded 
with respect to a parameter $\zeta$ as
\begin{eqnarray}
  \label{eq:partition}
  Z(\zeta)
  &=&
  e^{g(\zeta)}\prod_{n=1}^{\infty}(1-\frac{\zeta}{a_n})
  e^{\frac{\zeta}{a_n}+\frac{1}{2}(\frac{\zeta}{a_n})^2
    +\cdots+\frac{1}{m_n}(\frac{\zeta}{a_n})^{m_n}}, 
\end{eqnarray}
where the ${a_n}$ are not on the real axis.  Here, 
$\zeta$ may be the inverse of the temperature $\beta$, 
an external field, or any other parameters of the Hamiltonian.  
In most cases, partition functions are physically relevant 
only for real values of $\zeta$.  
Eq. (\ref{eq:partition}) yields an expression for 
the free energy, 
\begin{eqnarray}
  &&F(\zeta)   = 
  -\frac{C}{\beta}\log{Z(\zeta)}\cr
  &&=
  -\frac{C}{\beta}
  \left\{
  g(\zeta)
  +\sum_{n=1}^{\infty}\log{(1-\frac{\zeta}{a_n})}
  +\frac{\zeta}{a_n}+\frac{1}{2}(\frac{\zeta}{a_n})^2+\cdots+\frac{1}{m_n}(\frac{\zeta}{a_n})^{m_n}
  \right\}
\label{free}
\end{eqnarray}
where $\beta>0$ denotes the inverse temperature and
$C$ is a constant with respect to $\zeta$, introduced 
so that the thermodynamic limit is well-defined.  
For systems of finite size, $F(\zeta)$ is analytic in a neighborhood of the 
real axis since the $a_n$ are located a finite distance away from the real axis. Therefore, systems of 
finite size do not exhibit phase transitions 
with respect to $\zeta$. 

In the thermodynamic limit, however, the $a_n$ can become dense and 
may approach the real axis.  In such cases, the summation of 
the logarithm in eq. (\ref{free}) is replaced by an integral 
with an integral contour which intersects the real axis.  
As a consequence, the free energy is expressed by 
a different analytic function on each side of the 
integral contour.  This implies that a phase 
transition occurs at the intersection point 
as $\zeta$ varies along the real axis.

\subsection{Simple example}
\label{sec:example}
To illustrate the above scenario intuitively, 
we consider here a simple example.  
Let us suppose that the partition function 
of the example model can be written as, 
\begin{eqnarray}
  \label{eq:cosh}
  Z(\zeta;a,\zeta_c)
  =
  \cosh{\frac{\pi}{2 a}(\zeta-\zeta_c)}
\end{eqnarray}
where $a$ and $\zeta_c$ are real positive numbers.  The free energy is then
\begin{eqnarray}
  \label{eq:freedef}
  F(\zeta;a,\zeta_c)
  =
  -\frac{2a}{\beta}
  \log{Z(\zeta;a,\zeta_c)}.
\end{eqnarray}
where $2a$ is introduced to make the thermodynamic limit 
of this model well-defined.  
Since $\cosh{z} \simeq e^{{\rm sign}({\rm Re}(z)) z}/2$ 
for $|z| \gg 1$ holds , $Z(\zeta;a,\zeta_c)$ 
behaves like 
\begin{eqnarray}
  \label{eq:zlimit}
&&  Z(\zeta;a \to 0,\zeta_c) \cr
&&  \rightarrow
  \frac{1}{2}e^{\frac{\pi}{2a}(\zeta-\zeta_c)}\Theta({\rm Re}(\zeta-\zeta_c))
  +
  \frac{1}{2}e^{-\frac{\pi}{2a}(\zeta-\zeta_c)}\Theta(-{\rm Re}
(\zeta-\zeta_c)), 
\end{eqnarray}
in the limit as $a \rightarrow 0$, 
where ${\rm sign}(x)=x/|x|$ for $x \ne 0$, and $\Theta (x)=1$ 
for $x>0$ and vanishes for $x<0$. 
${\rm Re}(z)$ denotes the real part of a complex number $z$. 
This shows that the free energy can be expressed as 
\begin{eqnarray}
  \label{flimit}
  F(\zeta;a \to 0,\zeta_c)
  =
  -\frac{\pi }{\beta}(\zeta-\zeta_c)
  \{
  \Theta({\rm Re}(\zeta-\zeta_c))-\Theta(-{\rm Re}(\zeta-\zeta_c))
  \}, 
\label{first_order_transition} 
\end{eqnarray}
in this limit. 
This means that the analytic property of the free energy 
changes at the boundary ${\rm Re}(\zeta)=\zeta_c$;
more precisely, the first derivative of 
$F(\zeta;a,\zeta_c)$ with respect to $\zeta$ becomes 
discontinuous at ${\rm Re}(\zeta)=\zeta_c$
whereas the real part of the free energy remains continuous. 
If the parameter $\zeta$ is the temperature or an 
external field, this indicates that 
a first order phase transition occurs at $\zeta=\zeta_c$.

\begin{figure}
       \centerline{\includegraphics[width=8cm]
                                   {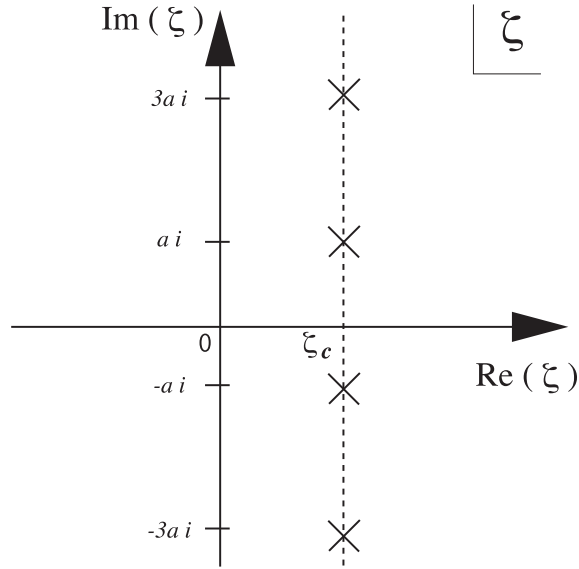}}
\caption{Zero points of the partition function $ Z(\zeta;a,\zeta_c)
  =
  \cosh{\frac{\pi}{2 a}(\zeta-\zeta_c)}$.  They become dense and approach the real axis in the limit as $a \rightarrow 0$}
\label{samplezero}
\end{figure}

This transition can be linked to the
asymptotic behavior of zeros of $\zeta$ in the complex plane, 
which arises in the ``artificial thermodynamic limit'' 
as $a \to 0$, as follows. 
Using the product expansion of $\cosh{z}$,
\begin{eqnarray}
  \label{eq:coshz}
  \cosh{z}
  =
  \prod_{k=1}^{\infty}
  \left(
    1+\frac{4z^2}{(2k-1)^2\pi^2}
  \right),
\end{eqnarray}
we can write
\begin{eqnarray}
  \label{eq:zexpand}
  Z(\zeta;a,\zeta_c)
  =
  \prod_{k=1}^{\infty}
  \left(
    1+\frac{(\zeta-\zeta_c)^2}{(2k-1)^2a^2}
  \right), 
\label{coshzeros}
\end{eqnarray}
which implies that $g(\zeta)$ and $m_n$ are zero 
in the expression of eq.(\ref{eq:partition}).  
Eq. (\ref{coshzeros}) means that 
zeros of $Z(\zeta;a,\zeta_c)$ do not occur on the real axis
but are distributed on $\zeta=\zeta_c\pm (2k-1)a \img$ 
$(k=1,2, \ldots)$ as is depicted in figure \ref{samplezero}, 
where $\img=\sqrt{-1}$. 
Therefore, the free energy of this model can be expressed as 
\begin{eqnarray}
  \label{eq:free}
    F(\zeta;a,\zeta_c)
  =
  -\frac{2a}{\beta}
  \sum_{k=1}^{\infty}\log{
  \left(
    1+\frac{(\zeta-\zeta_c)^2}{(2k-1)^2a^2}
  \right)}.
\end{eqnarray}
All the zeros $\zeta=\zeta_c\pm (2k-1)a \img$ correspond 
to branch points of the logarithm. 
Nevertheless, the free energy is still analytic along the real axis 
as long as $a$ is finite since there are no branch points on the real axis.

However, the situation changes in the limit as $a \to 0$
since the distance of zeros from the real axis becomes 
infinitesimal. Actually, the summation has to be 
replaced by an integral thus: 
\begin{eqnarray}
  \label{eq:integral}
      F(\zeta;a\to 0,\zeta_c)
  =
  -\frac{1}{\beta}
  \int_0^{\infty}dx
  \log{
  \left(
    1+\frac{(\zeta-\zeta_c)^2}{x^2}
  \right)},
\label{continuous_limit}
\end{eqnarray}
which implies that the integrand becomes singular
at $x=|\zeta-\zeta_c|$ when $\img(\zeta-\zeta_c)$ 
or $-\img(\zeta-\zeta_c)$ is a non-negatie real number. 
As a consequence, 
this integral correctly reproduces
eq. (\ref{first_order_transition}), which is singular
along a line ${\rm C1}:\zeta=\zeta_c+ v \img $ parameterized by $v \in \mR$
including the real critical parameter $\zeta=\zeta_c$. 

\subsection{Remarks}
Two issues are noteworthy here. 
\subsubsection{Argument principle and the number of zeros}
The first issue concerns the number of zeros inside a
closed contour. For any complex function $f(z)$ which is meromorphic inside
a closed contour $\gamma$, the identity
\begin{eqnarray}
\frac{1}{2 \pi \img} \int_{\gamma}
\frac{f^\prime (z)}{f(z)} dz=\sum_{j} n(\gamma,a_j)-\sum_{k}n(\gamma,b_k), 
\label{argment_principle}
\end{eqnarray}
holds, where $a_j$ and $b_k$ are the zeros and poles inside $\gamma$, 
respectively. 
$n(\gamma,p)$ represents the winding number 
of $\gamma$ around $p$. 
This formula is sometimes termed the {\em argument principle} \cite{Alf}. 
Applying this to eqs. (\ref{eq:partition}) and 
(\ref{free}), one can evaluate the number of zeros 
in the region surrounded by $\gamma$ 
\footnote{Eq. (\ref{argment_principle})
is analogous to Gauss's law for a two dimensional electromagnetic field. 
In this analogy, the free energy and zeros (poles) 
correspond to the electrostatic potential and point charges, 
respectively. 
}. 
In most cases, analytically finding all zeros is difficult
and one has to resort to numerical schemes such as Newton's method. 
A major difficulty of such approaches is to determine whether there has been
a  sufficient number of search trials. 
On the other hand, in some cases, the free energy density can be 
obtained in a computationally feasible manner in 
the thermodynamic limit. Therefore, one can estimate 
the asymptotic number of zeros by applying 
eq. (\ref{argment_principle}) to the free energy density. 
This approach can be utilized to check whether or not sufficiently 
many zeros have been obtained. 

\subsubsection{Incident angle and type of phase transition}
The second issue is the relation between the locus of the zeros 
and the type of the phase transition. 
In the simple model mentioned above, the locus that the zeros form in 
the limit as $a \to 0$ is the straight line ${\rm C1}:\zeta=\zeta_c+v \img$
$(v \in \mR)$, which is perpendicular to the real axis. 
This can be reproduced by dealing only with the limiting form of free energy 
(\ref{first_order_transition}) as follows. 
Eq. (\ref{first_order_transition}) means
that in the limit as $a \to 0$, the 
free energy is expressed by either of two analytic functions 
$F_{\rm 1}(\zeta)= -(\pi/\beta)(\zeta-\zeta_c)$ 
and 
$F_{\rm 2}(\zeta)= (\pi/\beta)(\zeta-\zeta_c)$, depending 
on the region to which $\zeta$ belongs. 
The critical condition for the selection 
is provided by 
${\rm Re}\left ( F_{\rm 1}(\zeta)\right )
={\rm Re}\left ( F_{\rm 2}(\zeta)\right )$, 
which yields the locus ${\rm C1}$. 
Such an argument can be generalized to some extent. 
Let us suppose that the free energy in the thermodynamic limit 
can be expressed by either of two analytic solutions 
$F_{\rm 1}(\zeta)$ and $F_{\rm 2}(\zeta)$ depending on 
$\zeta$, and the first discontinuity appears at the level of 
the first derivative. This means that the two analytic solutions are 
expanded as
\begin{eqnarray}
F_{\rm 1}(\zeta)=-\frac{1}{\beta} \left (A_0+A_1 (\zeta-\zeta_c)+
A_2 (\zeta-\zeta_c)^2 + \ldots \right ), 
\label{F1expand}
\end{eqnarray}
and 
\begin{eqnarray}
F_{\rm 2}(\zeta)=-\frac{1}{\beta}\left (B_0+B_1 (\zeta-\zeta_c)+
B_2 (\zeta-\zeta_c)^2 + \ldots \right ), 
\label{F2expand}
\end{eqnarray}
where $A_0=B_0$ and $A_1 \ne B_1$. 
A simple scenario implies that when the scale factor 
$a$ is small but finite, the partition function can be asymptotically 
expressed as
\begin{eqnarray}
&&Z(\zeta;a,\zeta_c) \simeq 
p_1 \exp \left (-\frac{\beta F_{\rm 1}(\zeta)}{a} \right ) + 
p_2 \exp \left (-\frac{\beta F_{\rm 2} (\zeta)}{a} \right ) \cr
&&\propto  \exp \left (\frac{A_0}{a}+
\frac{(A_1+B_1)(\zeta-\zeta_c)}{2a} \right ) 
\times \cosh \left (\frac{(A_1-B_1)(\zeta-\zeta_c)}{2a}+\theta \right ), 
\label{Fasymptotic}
\end{eqnarray}
for $|\zeta-\zeta_c| \ll a^{1/2}$, 
where $p_1$ and $p_2$ are prefactors
subexponential with respect to $a^{-1}$ 
and $ |\theta | \ll a^{-1}$. 
Since $\exp \left (A_0/a+  
(A_1+B_1)(\zeta-\zeta_c)/(2a) \right )$ never vanishes, 
zeros in the vicinity of $\zeta_c$, satisfying $|\zeta-\zeta_c| \ll a^{1/2}$, 
come out from only the cosh part, and are expressed as
$\zeta_k \simeq \zeta_c 
\pm (2 k-1)\pi a/(A_1-B_1) \img -2 a \theta /(A_1-B_1)$ 
($k=1,2,\ldots$).
In the limit of $a \to +0$, distances of contiguous zeros become 
infinitesimal and the locus of the zeros is parameterized as 
$\zeta \simeq \zeta_c +v \img$ ($v \in \mR$) as $a|\theta| \to 0$. 

This argument indicates that the condition 
${\rm Re}\left (F_{\rm 1}(\zeta) \right )=
{\rm Re}\left (F_{\rm 2}(\zeta) \right )$ forms a locus
of zeros in the thermodynamic limit and, as long as the transition 
is of the first order, 
which means that the first derivative of the free energy 
becomes discontinuous at a critical parameter value $\zeta_c$, 
the locus makes an incident angle (defined in the upper-half 
plane hereafter) $\pi/2$ to the real axis. 
The angle, however, depends on the type of the transition 
 \cite{Gro1,Gro2,Marinari1984}. 
Let us suppose a case of the second order 
phase transition, which corresponds to a situation 
$A_0=B_0$, $A_1=B_1$ and $A_2 \ne B_2$ in the above setting. 
For small but finite $a$, an argument similar to the above yields an
expression 
\begin{eqnarray}
Z(\zeta;a,\zeta_c) 
&\propto&  \exp \left (\frac{A_0}{a}+
\frac{A_1(\zeta-\zeta_c)}{a} 
+\frac{(A_2+B_2)(\zeta-\zeta_c)^2}{2a}\right )  \cr
&& \phantom{\exp} 
\times \cosh \left (\frac{(A_2-B_2)(\zeta-\zeta_c)^2}{2a}+\theta \right ), 
\label{Fasymptotic2}
\end{eqnarray}
for $|\zeta-\zeta_c| \ll a^{1/3}$, implying that zeros in the vicinity of 
$\zeta_c$ are asymptotically expressed as
$\zeta \simeq \zeta_c +v \exp \left (
\pm (2l-1)\pi/4 \img \right )$ $(l=1, 2, \ v \in \mR)$ and 
the incident angle to the real axis is not $\pi/2$ but 
$\pi/4$ and $3\pi/4$. 
More generally, when the first discontinuity of the free energy comes out 
at the $m$th derivative, possible incident angles are limited to 
forms of $(2l-1)\pi/(2m)$ ($l=1,2,\ldots,m$) \cite{Kenna2006}. 
In this way, the profile of the locus provides
us with a useful clue for classifying the types 
of phase transition. 

Notice that the above argument crucially relies on 
an assumption that the subexponential prefactors 
$p_1$ and $p_2$ do not vanish in the vicinity of $\zeta_c$. 
In some cases, however, either of them can vanish in the 
left or right side of $\zeta_c$, 
which excludes some of the multiple branches of the incident angles. 
Later, we will see that this does occur in DREM. 

\section{Complex zeros with respect to the replica number 
for a discrete random energy model}
\label{sec:replica}
Now, we are ready to employ the Lee-Yang type approach
for characterizing analyticity breaking with respect to 
the replica number which occurs in DREM.
\subsection{Model definition and an exact expression of the 
moment of the partition function}
A DREM is defined by 
sampling $2^N$ energy states 
$\epsilon_1, \epsilon_2, \ldots, \epsilon_{2^N}$
independently from an identical distribution 
\begin{eqnarray}
P(E_i)=2^{-M} \left (
\begin{array}{c}
M \cr
\frac{1}{2}M+E_i
\end{array}
\right ), 
\label{DREM_dist}
\end{eqnarray}
where $N$ and $M=\alpha N$ are positive integers, 
and $E_i=i-M/2$ for $i=0,1,\ldots,M$. 

\begin{figure}
       \centerline{\includegraphics[width=8cm]
                                   {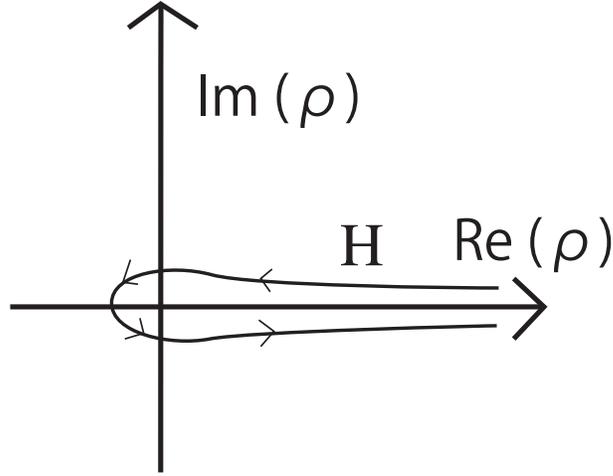}}
\caption{Integration contour to calculate Eq.(\ref{exactmoment}).}
\label{contor}
\end{figure}

For a given realization of energy levels $\{\epsilon_i\}$, 
the partition function of an inverse temperature 
$\beta$ is defined as
$Z=\sum_{i=1}^{2^N}\exp \left (-\beta \epsilon_i \right )$. 
In \cite{KO2}, the authors showed that for $\forall{n} \in \mC$, 
$\forall{N}$ and $\forall{M}$ the moment of the 
partition function can be expressed as
\begin{eqnarray}
\left<Z^n\right>
&=&
\frac{
\omega^{-\frac{nM}{2}}
}{
\tilde\Gamma (-n)
}
\int_H d\rho(-\rho)^{-n-1}
\left(\sum_{i=0}^M P(E_i)e^{-\omega^i\rho}\right)^{2^N}
\label{exactmoment}
\end{eqnarray}
where 
$\omega=e^{-\beta}$, 
$\tilde\Gamma(n)\equiv -2 \img \sin (n\pi) \Gamma(n)$ and 
the integration contour $H$ is defined as shown in figure \ref{contor}.
$\Gamma(z)$ is the Gamma function. 
From this expression, one can analytically determine
the limit $\phi(n)=\lim_{N \to \infty} N^{-1}\log \left<Z^n\right>$, 
which indicates the following behavior. 
Let us define a critical inverse temperature as
\begin{eqnarray}
\beta_c=\left \{
\begin{array}{ll}
\infty, & \alpha \le 1, \cr
\log\left(1-h_2^{-1}(1-\alpha^{-1}
)\right )-
\log\left(h_2^{-1}(1-\alpha^{-1}
)\right ), & \alpha > 1, 
\end{array}
\right .
\end{eqnarray}
where $h_2^{-1}(y)$
is the inverse function of the binary 
entropy $h_2(x)=-x \log_2(x)-(1-x) \log_2(1-x)$
for $0<x<1/2$. 
For $\beta < \beta_c$, $\phi(n)$ is provided by either of 
\begin{eqnarray}
\phi_{\rm RS1}(n)=n \left ( \log 2+ \alpha \log \left (
\cosh \left( \frac{\beta}{2} \right ) \right ) \right ), 
\label{rs1}
\end{eqnarray}
or 
\begin{eqnarray}
\phi_{\rm RS2}(n)=\left ( \log 2+ \alpha \log \left (
\cosh \left( \frac{n\beta}{2} \right ) \right ) \right ), 
\label{rs2}
\end{eqnarray}
depending on $n$. More precisely, there exists $\exists{n}_{\rm RS}>1$
such that $\phi(n)=\phi_{\rm RS2}(n)$ for $n > n_{\rm RS}$ whereas
$\phi(n)=\phi_{\rm RS1}(n)$ for $n \le n_{\rm RS}$. 
On the other hand, the behavior is different for $\beta > \beta_c$. 
For $n> n_c=\beta_c/\beta $, $\phi(n)=\phi_{\rm RS2}(n)$ holds. 
However, $\phi(n)$ for $n < n_c$ is described by 
neither $\phi_{\rm RS1}(n)$ nor $\phi_{\rm RS2}(n)$
but another solution 
\begin{eqnarray}
\phi_{\rm 1RSB}(n)=\frac{n\alpha\beta}{2} \tanh\frac{\beta_c}{2}. 
\label{1rsb}
\end{eqnarray}

Comparison with the replica analysis indicates 
$\phi_{\rm 1RSB}(n)$ corresponds to the one step replica symmetric 
(1RSB) solution \cite{Mou1,Mou2}. 
For DREM, the uniqueness of analytical continuation 
from $n \in \mN$ to $n \in \mC$ is guaranteed by Carlson's theorem 
\cite{TIT}. 
This indicates that the analyticity breaking at $n=n_c$ for $\beta > \beta_c$
is the origin of the 1RSB transition. 

In addition to the advantage of yielding analytical expressions 
(\ref{rs1}), (\ref{rs2}) and (\ref{1rsb}) in the thermodynamic
limit as $N \to \infty$, eq. (\ref{exactmoment}) is useful for numerically 
evaluating $\left \langle Z^n \right \rangle$ for 
finite $N$ because the necessary cost for the computation 
grows only linearly with $N$. 
This property holds for $\forall{n} \in \mC$,  
which is advantageous in searching for zeros 
of $\left \langle Z^n \right \rangle$ with respect to 
$n \in \mC$ in order to characterize transitions 
that occur in the limit as $N \to \infty$. 

\subsection{Analytical results}
Let us apply the argument of the previous section to DREM. 
For simplicity, we focus here on the case of 
the 1RSB transition at $n=n_c$ assuming 
$\alpha > 1$ and $\beta > \beta_c$. 

\subsubsection{Locus of zeros}
Eqs. (\ref{rs2}) and (\ref{1rsb}) indicate that 
$\left (\partial^2/\partial n^2 \right ) \phi_{\rm RS2}(n)
=(\alpha \beta^2/4)\left (1-\tanh^2(\beta_c/2) \right )
> \left (\partial^2/\partial n^2 \right ) \phi_{\rm 1RSB}(n)=0$
at  $n=n_c=\beta_c/\beta$, 
whereas $\left (\partial /\partial n \right ) \phi_{\rm RS2}(n)
=\left (\partial /\partial n \right ) \phi_{\rm 1RSB}(n)
=(\alpha \beta/2) \tanh(\beta_c/2)$ holds.
This implies that the 1RSB transition is classified to 
the second order. The naive argument provided in the previous
section means that the locus of zeros can be locally parameterized as
\begin{eqnarray}
n\simeq n_c +v \exp\left (\pm (2l-1)\pi/4 \img \right ), 
\label{locus}
\end{eqnarray}
where $l=1,2$, $v \in \mR$. However, this result must be corrected; 
the branch of $l=2$, which corresponds to the incident 
angle $3\pi/4$, never appears due to the following reason. 

For $N \gg 1$, asymptotic evaluation of eq. 
(\ref{exactmoment}) yields an expression 
\begin{eqnarray}
\left \langle Z^n \right \rangle 
&\simeq& N \alpha \beta e^{Nn \alpha \beta/2} \cr
&\times&\int_0^{h_2^{-1}(1-\alpha^{-1})} dy 
\exp \left (N \left (-n \alpha \beta y+
\left (1-\alpha + \alpha h_2(y) \right )\log 2 \right ) \right ), 
\label{gumbel_expression}
\end{eqnarray}
derivation of which is shown in \cite{KO2}.	
Evaluating dominant contributions in the integral by 
the steepest descent method \cite{Keener}
indicates that the asymptotic expression 
is further simplified as
\begin{eqnarray}
\left \langle Z^n \right \rangle \simeq 
\left \{
\begin{array}{ll}
p_{\rm RS2} e^{N \phi_{\rm RS2}(n) }
+
p_{\rm 1RSB} e^{N \phi_{\rm 1RSB}(n) }, & n_c < {\rm Re}(n) < 1, \cr
p_{\rm 1RSB} e^{N \phi_{\rm 1RSB}(n) }, &0< {\rm Re}(n) < n_c, 
\end{array}
\right . 
\label{correct_asymptotics}
\end{eqnarray}
depending on the position of $n$, 
where $p_{\rm RS2}=(\pi/(2N\alpha))^{1/2} (\cosh (n\beta/2))^{-1}$
and $p_{\rm 1RSB}=(N\alpha \beta(n_c-n))^{-1}$. 
This is because the absolute value of 
the integrand is maximized in the integral interval 
$0 < y < h_2^{-1}(1-\alpha^{-1})$ for $n_c < {\rm Re}(n) < 1$
while the right terminal point $y=h_2^{-1}(1-\alpha^{-1})$ offers a
unique dominant contribution for $0<{\rm Re}(n) < n_c$. 
Equation (\ref{correct_asymptotics}) means that, 
for $N \gg 1$, there are no zeros of  $0<{\rm Re}(n) < n_c$.
Solving $\left \langle Z^n \right \rangle =0$
with respect to $n$ perturbatively 
in the vicinity of $n =n_c$ for $n_c<{\rm Re}(n)<1$ 
under a condition of $N \gg 1$ yields
an asymptotic expression of the zeros 
\begin{eqnarray}
&&n_k\simeq n_c +\frac{4}{\beta}\left (1-\frac{1}{16k} \right )
\cosh \left (\frac{\beta_c}{2} \right )
\sqrt{\frac{k \pi}{N \alpha}}  \cr
&& \phantom{n\simeq n_c +} \times 
\exp \left (\pm  
\left (\frac{\pi}{4}+\frac{1}{8k \pi}\log (8k\pi^2) \right )\img \right )
+ o\left (N^{-1/2} \right ), 
\label{locus_correct}
\end{eqnarray}
$(k=1,2,\ldots)$, where $o\left (N^{-1/2} \right )$ represents
contributions which are relatively negligible compared to $N^{-1/2}$. 

This result indicates that the distance from 
the real axis to the closest zero, $n_1$, 
decays as $O(N^{-1/2})$ and, 
in the thermodynamic limit $N \to \infty$, 
the incident angle of the phase boundary converges to $\pi/4$. 
However, the angle between the real axis and
the line connecting the transition point $n_c$ 
with the $k$th zero, $n_k$ $(k=1,2,\ldots)$, 
is larger than $\pi/4$ by $1/(8 k\pi)\log (8k\pi^2)$ 
independently of $N$ as long as $k$ is finite.	
This implies that for systems of finite sizes zeros in the vicinity of 
$n_c$ are expected to be placed in the left side of the phase boundary. 

\begin{figure}
       \centerline{\includegraphics[width=8cm]
                                   {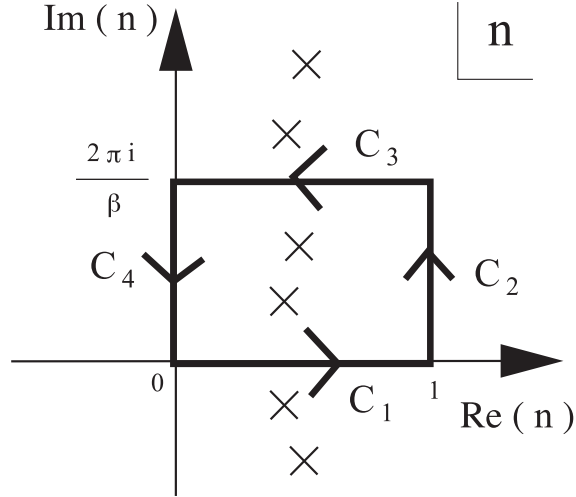}}
\caption{A cycle $\gamma$, the number of 
complex zeros inside which is evaluated utilizing 
the argument principle. }
\label{loop}
\end{figure}

\subsubsection{Number of zeros}
Eqs. (\ref{rs2}) and (\ref{1rsb}) can also be utilized 
to evaluate the asymptotic number of complex zeros. 
For this, let us consider a closed cycle 
$\gamma: C_1 \to C_2 \to C_3 \to C_4$ as shown in figure \ref{loop}. 
Application of the argument principle of the 
previous section to eqs. (\ref{rs2}) and (\ref{1rsb})
yields the asymptotic number of zeros 
inside $\gamma$. $\gamma$ in figure \ref{loop} guarantees that 
the estimate does not vanish since the critical 
replica number $n_c=\beta_c/\beta$ is in the range $0 < n < 1$. 

To simplify the analysis, 
we decompose the generating function as 
\begin{eqnarray}
  \label{eq:poleexpression}
\phi_{N}(n)=\frac{1}{N}\log  \left<Z^n\right>
  = \Lambda(n) +\img \Omega (n)
\end{eqnarray}
where $\Lambda(n)$ and $\Omega (n)$ are real functions.  
The argument principle indicates that 
$N (2 \pi )^{-1} \int_{\gamma} dn (\partial /\partial n)\Omega (n)$, 
which is proportional to the total variation of 
$\Omega(n)$ as $n$ goes around $\gamma$, 
accords with the number of zeros inside $\gamma$. 
For $N \gg 1$, we substitute $\phi(n)$ into $\phi_N(n)$, 
which makes it possible to analytically evaluate 
the variation. 

Along $C_1$, both $\phi_{\rm RS2}(n)$ and $\phi_{\rm 1RSB}(n)$ 
lead to $\Omega(n)=0$. 
Therefore, 
\begin{eqnarray}
  \label{eq:c1}
  \Delta_{C_1} \Omega=\Omega(1)-\Omega(0)=0. 
\end{eqnarray}
Along $C_2$, $\phi(n)=\phi_{\rm RS2}(n)=\log(2)+\alpha \log \left (
\cosh \left (n \beta/2 \right ) \right )$ holds. 
This leads to 
\begin{eqnarray}
  \label{eq:c2delta}
    \Delta_{C_2} \Omega 
    &=&
{\rm Im}\left (\phi_{\rm RS2}\left (1+(2 \pi/\beta)\img  \right ) \right )
-{\rm Im}\left (\phi_{\rm RS2}\left (1  \right ) \right ) \cr
&=&{\rm Im}\left (\alpha \log\left (-\cosh(\beta/2)   \right ) \right )
=\alpha \pi, 
\end{eqnarray}
where ${\rm Im}(z)$ denotes the imaginary part of a complex number $z$. 
Along $C_3$, $\phi(n)$ becomes real and therefore
\begin{eqnarray}
\Delta_{C_3} \Omega=0. 
\label{eq:c3}
\end{eqnarray}
Along $C_4$, on the other hand, $\phi(n)=\phi_{\rm 1RSB}(n)$ holds, 
which yields
\begin{eqnarray}
  \label{eq:c4delta}
    \Delta_{C_4} \Omega 
    &=&{\rm Im}\left (\phi_{\rm 1RSB}\left (0\right ) \right )
-{\rm Im}\left (\phi_{\rm 1RBS}\left ((2 \pi/\beta) \img  \right ) \right ) \cr
&=&-{\rm Im}\left ( \frac{\alpha \beta }{2} \tanh \left (
\frac{\beta_c}{2} \right ) \frac{2 \pi \img }{\beta} \right )
=-{\alpha \pi} \tanh \left (\frac{\beta_c}{2}\right ). 
\end{eqnarray}
These results indicate that the total number of zeros inside 
the cycle in the low temperature phase $\beta > \beta_c$
can be asymptotically estimated as 
\begin{eqnarray}
  \label{eq:nu}
  \nu_{est}
  \simeq \frac{N}{2 \pi} \left (
\Delta_{C_1}\Omega+\Delta_{C_2}\Omega
+\Delta_{C_3}\Omega+\Delta_{C_4}\Omega \right )
= \frac{\alpha N}{1+e^{\beta_c}}
\end{eqnarray}
for $N \gg 1$. 

This indicates that the number of the zeros
inside $\gamma$ grows proportionally to $N$. 
However, the total number of zeros over the complex plane 
of $n$ is infinite because
$\left \langle Z^n \right \rangle$ is a transcendent function 
of $n$ and has period $2 \pi/\beta$ in the imaginary direction. 
This is in contrast to the case of the Lee-Yang zeros with respect 
to $\beta$ for a typical sample of DREM \cite{Mou1,Mou2}.
In this case, the partition function $Z(\beta)$ becomes a polynomial 
of $\beta$ and the number of zeros over the entire plane
grows only linearly with respect to $N$, 
which yields a continuous distribution of zeros
after being averaged over $\{\epsilon_i\}$.

\subsection{Numerical results}
In order to justify the analytically obtained results for $N \gg 1$, 
we numerically examined the zeros of $\left \langle Z^n \right 
\rangle =0$ for several $N$ utilizing the expression 
of eq. (\ref{exactmoment}). Unfortunately, as eq. (\ref{exactmoment}) 
is transcendent with respect to $n$, 
solving for the zeros algebraically is difficult. 
Therefore, we resorted to an iterative numerical method to  
search for the zeros. 

We employed the {\em secant method} \cite{secant}. 
For solving an equation $f(z)=0$, 
this scheme iterates the recurrence relation 
\begin{eqnarray}
z_{t+1}=z_t-
\frac{z_t-z_{t-1}}{
f(z_t)-f(z_{t-1})}f(z_t), 
\label{secant}
\end{eqnarray}
until $|z_{t}-z_{t-1}|$ becomes smaller than 
a feasible discrepancy level, which was 
here chosen to be $10^{-8}$. 
This update rule is somewhat similar to that of Newton's method. 
Actually, the secant method can be regarded as an 
approximation of Newton's method, which is obtained by 
replacing $(z_t-z_{t-1})/(f(z_t)-f(z_{t-1}))$ with 
$f^\prime(z_t)$ in eq. (\ref{secant}). 
A major advantage of this method is that there is no need 
to evaluate the first derivative. 
This property is useful when $f(z)$ is 
complicated, which is the case for the current 
objective system

To find all zeros inside $\gamma: C_1 \to C_2 \to 
C_3 \to C_4$, we adopted the following strategy. 
We first covered the region inside $\gamma$ with a mesh of a 
fixed size in order to determine a set of initial 
values $n_1$ and $n_2$, which are required 
when utilizing the secant method. 
Two adjacent points in the vertical direction 
on the mesh were taken as the initial points. 
In the iteration of eq. (\ref{secant}), 
a calculated new point may step out of the region.  
In such cases, we changed the pair and tried the search again. 
Once we found a root $a_1$, we replaced 
$\left\langle Z^n\right\rangle$
with $\left\langle Z^n\right\rangle/(n-a_1)$ and continued the calculation.  
After trying every pair of initial values, 
we reduced the mesh size and searched for the roots again. 
We finished these procedures when we were unable to find a new root.   

Table \ref{number_of_zeros} 
shows dependence of the 
number of 
zeros on the system size $N$. 
This indicates that the number of zeros obtained numerically
is reasonably consistent with the number obtained from analytical evaluation 
(\ref{eq:nu}), implying that the number of the search trials has 
been sufficiently many. 

\begin{table}
  \centering
  \begin{tabular}{|c|c|c|}
    \hline
    $N$ & $\nu_{est}$ & $\nu$\\
    \hline
    5 &  4.3  &  5 \\
    10 & 8.6 & 10 \\
    15 & 12.9 & 14 \\
    20 & 17.1 & 19 \\
    25 & 21.4 & 23 \\
    \hline
  \end{tabular}
  \caption{The number of complex zeros with respect to $n$ 
in the cycle shown in figure \ref{loop} for $\alpha = 4$ 
and $\beta=3 \beta_c$.
The estimated number 
is calculated from eq. (\ref{eq:nu}) and the actual number $\nu$ 
is evaluated by numerically searching for the zeros using
eq. (\ref{secant}). }
\label{number_of_zeros} 
\end{table}

\begin{figure}
       \centerline{\includegraphics[width=12cm]
                                   {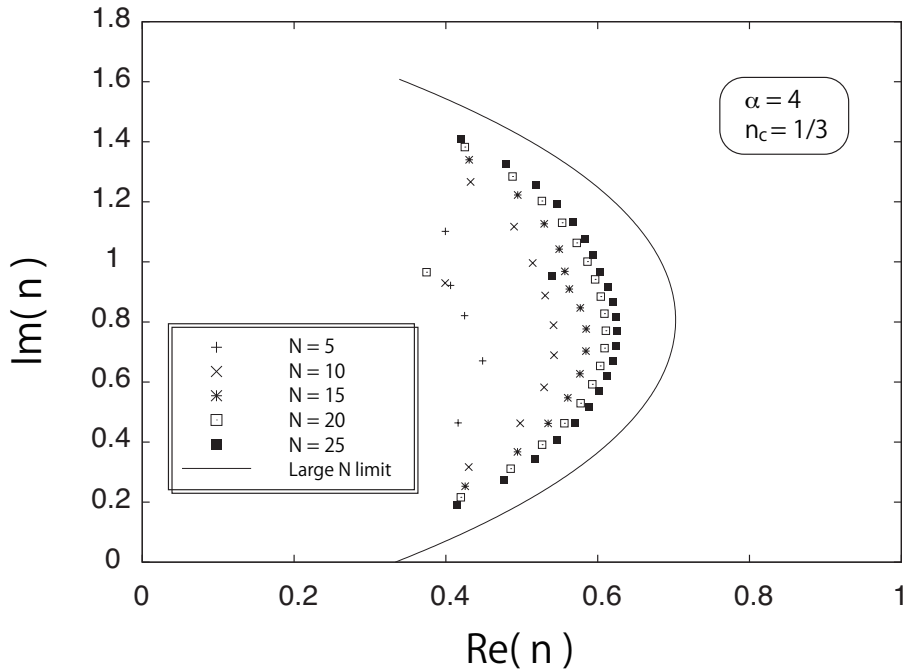}}
\caption{Complex zeros of the moment of the partition function 
of DREM for $\alpha = 4$ and $\beta = 3 \beta_c$.
The transition point in this case is determined to be $n_c=\beta_c/\beta=1/3$. 
The curve represents the locus of zeros in the thermodynamic 
limit, which is evaluated using ${\rm Re} (\phi_{\rm RS2}(n))=
{\rm Re} (\phi_{\rm 1RSB}(n))$.  
The incident angle of the curve to the real axis
is $\pi/4$, which indicates that the transition of $\phi(n)$
is analogous to a phase transition of the second order. 
Numerical data (markers) approach the locus as the system
size $N$ increases from $N=5$ to $25$ in steps of $5$.  }
\label{leeyangzeros}
\end{figure}

Figure \ref{leeyangzeros} shows results of the search for 
$N=5,10,15,20$ and $25$. 
The curve represents the locus of the zeros in the limit as 
$N \to \infty$, which was evaluated using the condition 
${\rm Re} (\phi_{\rm RS2}(n))={\rm Re} (\phi_{\rm 1RSB}(n))$. 
As we expected, the incident angle of the locus to
the real axis was $\pi/4$, indicating that 
the first discontinuity of $\phi(n)$ appears at 
the level of the second derivative. 

\begin{figure}
       \centerline{\includegraphics[width=12cm]
                                   {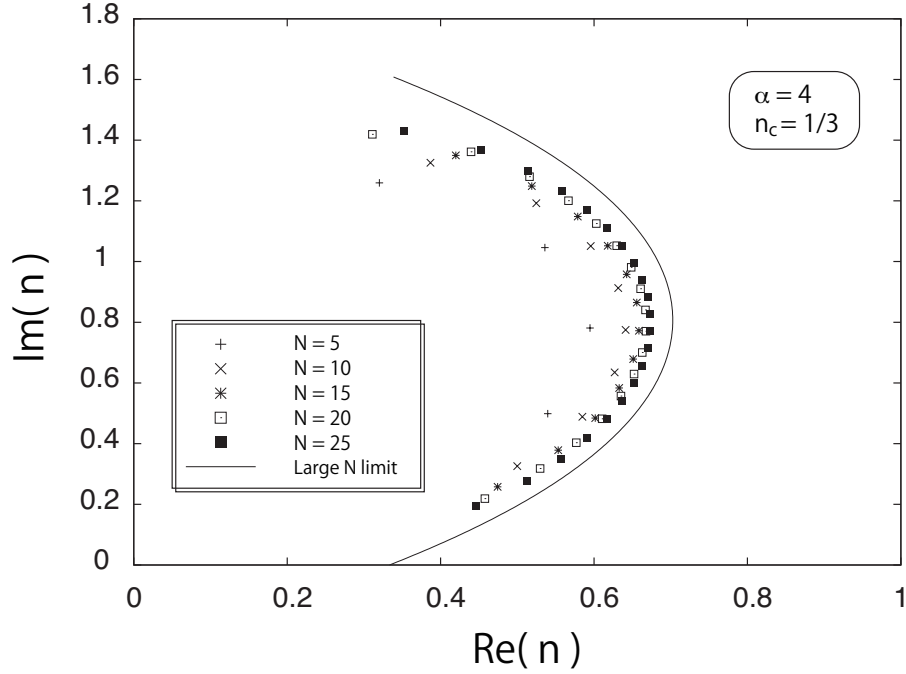}}
\caption{Complex zeros obtained from the asymptotic expression 
of eq. (\ref{correct_asymptotics}) for $\alpha = 4$, $\beta = 3 \beta_c$
and $N=5,10,\ldots,25$.
}
\label{leeyangzeros2}
\end{figure}

\begin{figure}
       \centerline{\includegraphics[width=12cm]
                                   {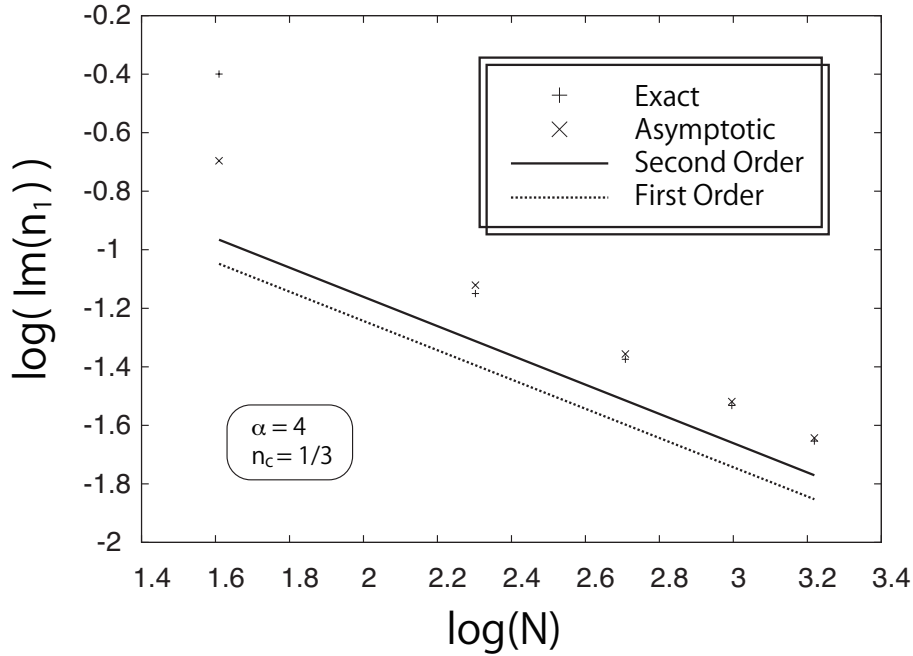}}
\caption{Distance of $n_1$ to the real axis (${\rm Im}(n_1)$)
versus system size $N$. 
Markers, $+$ and $\times$, represent the results 
for the exact expression of eq. (\ref{exactmoment}) 
and for the asymptotic one of eq. (\ref{correct_asymptotics}), 
respectively. The full line shows theoretical prediction
${\rm Im}(n_1)=5\sqrt{\pi}/(4\beta\sqrt{N\alpha})\cosh(\beta_c/2)
\sin(\pi/4+(1/8)\log(8 \pi^2))$, 
which is obtained by neglecting the $o(N^{-1/2})$ contribution 
in eq. (\ref{locus_correct}), while the correction 
$(1/8)\log(8 \pi^2)$ is omitted in the broken line. 
These support our theoretical predictions which indicate 
that the distance vanishes as $O(N^{-1/2})$ and positive corrections
of the incident angle from $\pi/4$ exist for $n_k$ independently of $N$ 
as long as $k=1,2,\ldots$ is finite. 
}
\label{asympdistance}
\end{figure}

The markers represent data obtained from the numerical 
search. These show that the zeros approach the locus 
from the left side as $N$ becomes larger, as we expected. 
Makers of figure \ref{leeyangzeros2} stand for zeros obtained
from the asymptotic expression of eq. (\ref{correct_asymptotics}). 
These exhibit behavior qualitatively similar to 
that of the exact expression; but deviation is not negligible, 
which is presumably due to finite size effects. 
Nevertheless, the distance of $n_1$ to the real axis (${\rm Im}(n_1)$)
shows a fairly good consistency with the theoretical prediction 
of eq. (\ref{locus_correct}), validating our asymptotic analysis
based on the steepest descent method (figure \ref{asympdistance}). 

In conclusion, the overall consistency between the analytical 
and numerical evaluations justifies the current 
analysis based on eq. (\ref{exactmoment}).

\section{Summary}\label{summary}
In summary, we have characterized analyticity breaking 
with respect to the replica number $n$ which arises in 
the discrete random energy model (DREM) by examining
zeros of the moment of the partition function 
over the complex $n$ plane. To do this, we utilized 
an exact expression of the moment of the partition function, 
which was introduced by the authors in a related paper \cite{KO2}. 
The expression is valid for $\forall{n} \in \mC$ and 
useful for numerically evaluating the moment of finite 
size systems. Taking the thermodynamic limit of the expression 
at sufficiently low temperatures indicates that 
analyticity breaking occurs at a critical
replica number $0<n_c < 1$, which can be regarded as
the origin of the one step RSB (1RSB) solution. 
Perturbatively evaluating the zeros in the vicinity of 
$n_c$ based on the expression utilizing the steepest descent method
shows that the distance from the real axis to the closest zero
decays as $O(N^{-1/2})$ when the system size $N$ is large. 
Examining the asymptotic form of the locus of the zeros
in the thermodynamic limit $N \to \infty$ 
implies that the transition is analogous to a phase transition 
of the second order.
We also evaluated the asymptotic number of zeros inside a 
unit cycle shown in figure \ref{loop} 
based on the expression. Zeros numerically obtained for 
finite size systems are reasonably consistent with 
the analytical predictions. 

The approach developed here is also applicable to the standard 
continuous random energy model \cite{REM}, and this approach yields results 
qualitatively the same as those for DREM. This implies that 
the 1RSB transitions observed in a family of REMs
can be generally characterized by the complex zeros in a similar manner. 

A natural question to ask is whether another type of RSB, 
full RSB (FRSB), can be characterized similarly by the complex zeros. 
Recently, one of the authors examined the zeros 
of tree systems in a vanishing temperature limit \cite{Obuchi2008}. 
Although some of the systems are conjectured to exhibit FRSB 
at a certain critical number, numerical data about the zeros
seem irrelevant to the FRSB transition. 
It is, therefore, desirable to explore other systems
in order to assess whether or not the irrelevance of the zeros 
to FRSB is specific to tree systems. 


\appendix

%

\ack
This work was supported by a Grant-in-Aid 
for Young Scientists (B) by JSPS (KO) 
and that on the Priority Area 
``Deepening and Expansion of Statistical Mechanical
Informatics'' by  the Ministry of Education, Culture, Sports and Science (YK).

\end{document}

%% file: commands.tex
\newcommand{\img}{{\rm i}}

\newcommand{\mC}{\mathbb{C}}
\newcommand{\mR}{\mathbb{R}}
\newcommand{\mN}{\mathbb{N}}